# Optical Properties of Bismuth Nanostructures Towards the Ultrathin Film Regime

Johann Toudert,[1,*] Rosalia Serna,[1,*] Claire Deeb,[2] Esther Rebollar[3]

[1]*Laser Processing Group, Instituto de Óptica, IO, CSIC, Serrano 121, 28006 Madrid, Spain*

[2]*Center for Nanoscience and Nanotechnology C2N, CNRS, Université Paris-Sud, Université Paris-Saclay, 10 Boulevard Thomas Gobert, 91120 Palaiseau, France*

[3]*Instituto de Química Física Rocasolano, IQFR, CSIC, Serrano 119, 28006 Madrid, Spain*

*\*e-mails: johann.toudert@gmail.com, rosalia.serna@csic.es*

**Abstract**. Bulk bismuth presents outstanding optical properties, such as a giant infrared refractive index (n ~ 10) and a negative ultraviolet - visible permittivity induced by giant interband electronic transitions. Although such properties are very appealing for applications in nanophotonics, the dielectric function of bismuth nanostructures has been scarcely studied. Here, we determine by spectroscopic ellipsometry the far infrared – to – ultraviolet dielectric function of pulsed laser deposited bismuth thin films with nominal thickness $t_{Bi}$ varied from near 10 nm to several tens of nm. For $t_{Bi} > 15$ nm, the films display a continuous structure and their dielectric function is comparable with that of bulk bismuth. For $t_{Bi} < 15$ nm, the film structure is discontinuous, and the dielectric function differs markedly from that of bulk bismuth. It is proposed from FDTD simulations that this marked difference arises mainly from effective medium effects induced by the discontinuous film structure, where quantum electronic confinement does not play a dominant role. This suggests that ultrathin and continuous bismuth films should present the same outstanding optical properties as bulk bismuth for high performance nanophotonic devices.

## 1. Introduction

Bulk bismuth (Bi) presents outstanding optical properties, related with its giant interband electronic transitions, such as a giant infrared refractive index (n ~ 10) and a negative ultraviolet – visible permittivity [1]. These properties are thought to enable a strong visible and infrared absorption [1-4] and an ultraviolet – visible plasmonic response [1-3, 5, 6] in deeply subwavelength Bi nanostructures. These effects are of utmost importance for a growing number of applications based on Bi nanostructures, including photocatalysis [7-10], photodetection [16,12], or optical modulation [13,14]. For exploiting the full potential of such applications by rational nanostructure design, knowing the dielectric function of Bi nanostructures in a broad spectral range, from the far infrared to the ultraviolet, is needed. However, such data are not available, despite of several claims of quantum electronic confinement effects implying a size-dependence for the electronic structure of bismuth nanostructures [15-18]. In fact, very few attempts to determine it were reported in the past years, in the case of bismuth thin films [19] and nanowires [15], and they were limited to narrow spectral windows in the infrared.

In here, we determine the far infrared – to – ultraviolet dielectric function of Bi thin films with nominal thicknesses $t_{Bi}$ varied from near 10 nm to several tens of nm. All the studied films display a continuous structure and a dielectric function comparable to that of bulk Bi, except the thinnest one ($t_{Bi}$ = 11 nm). In this case, our analysis suggests that the deviation from bulk values originates from the discontinuous film structure, the incident light interacting with an effective medium consisting of Bi and air. This leads us to propose that the observed deviation does not primarily originate from quantum electronic confinement.

## 2. Optical Properties of the Bi Films

The studied Bi films, grown on Si substrates, present nominal thicknesses $t_{Bi}$ = 11 nm, 17 nm, 21 nm, 28 nm, and 78 nm, determined by Rutherford backscattering spectroscopy (RBS). The optical properties of the films were characterized by spectroscopic ellipsometry at photon energies from the far infrared to the ultraviolet (0.05 to 4 eV, 25 to 0.3 μm). Details about the film fabrication and characterization, including ellipsometry measurements, are given in **Annex A1**. **Figure 1a** shows the experimental spectra of the ellipsometric angles Ψ and Δ for the studied films, measured at an angle of incidence of

70º. The full sets of spectra (measured at different angles of incidence) of these films are shown in **Annex A2** (**Figure S1**). Clear trends can be seen on both Ψ and Δ upon increasing $t_{Bi}$. For instance, Ψ increases in the whole far infrared – to – ultraviolet region.

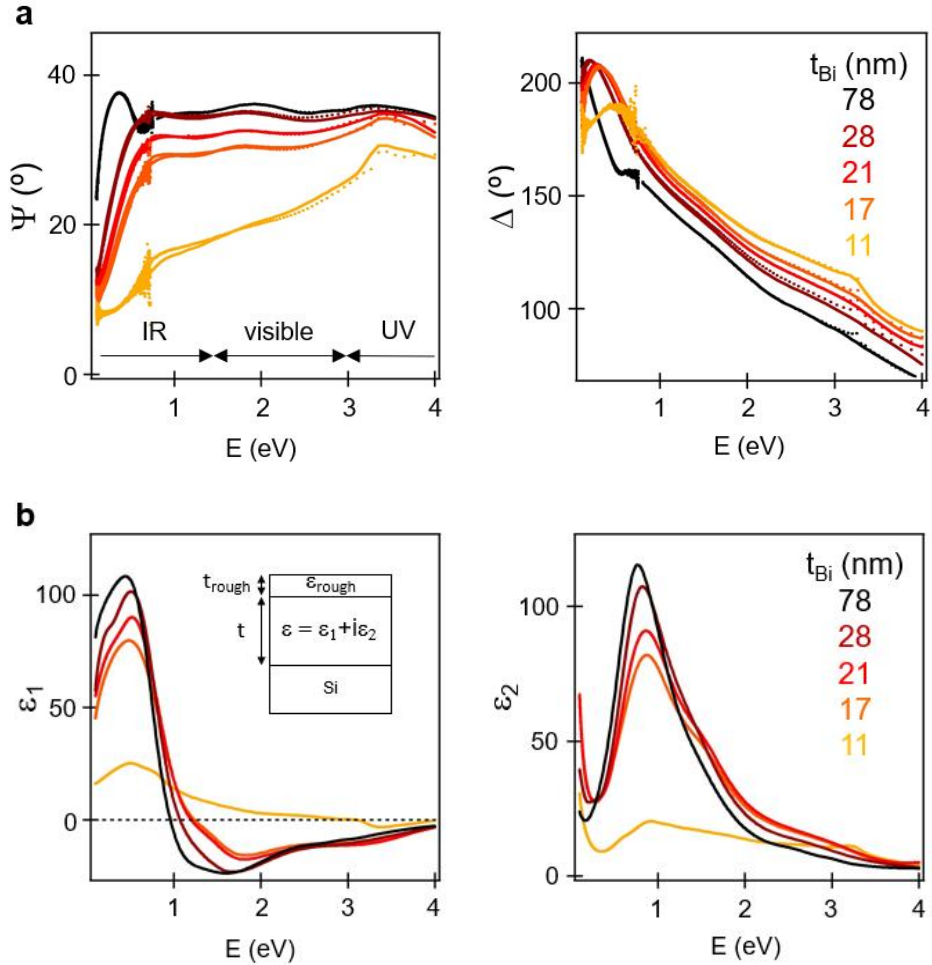

**Figure 1. Far infrared - to - ultraviolet ellipsometry characterization of the films and of their dielectric function.** (a) Spectra of the ellipsometric angles Ψ and Δ at the angle of incidence of 70º: experimental spectra (dots) and the corresponding best-fit ones (lines). (b) Spectra of the real and imaginary part ($\varepsilon_1$ and $\varepsilon_2$) of the best-fit dielectric function of each film. The model used for the ellipsometry analysis is shown on the left panel.

To determine the dielectric function $\varepsilon = \varepsilon_1 + i\varepsilon_2$ of each Bi film, its experimental Ψ and Δ spectra measured at different angles of incidence were fitted simultaneously using a bilayer model. This model, which is depicted in **Figure 1b** and further detailed in **Annex A2**, **Figure S2**, consisted of a layer of dielectric function $\varepsilon$ covered by a roughness layer of dielectric function $\varepsilon_{rough}$. $\varepsilon$ was described by a sum of Kramers Kronig – consistent oscillators. $\varepsilon_{rough}$ was described with a Bruggeman model mixing $\varepsilon$ and the dielectric function of air with an equal weight. The thicknesses of the two layers, t and $t_{rough}$ were fixed at the values found from the structural characterization, as explained in **Annex A1**. In short, $t_{rough}$ is the average peak-to-valley roughness found by atomic force microscopy (AFM), and t is the remaining film thickness underneath this roughness as determined by combining AFM, RBS and scanning electron microscopy (SEM). The t and $t_{rough}$ values are given in **Annex A2**, **Table S1**, together with the corresponding geometric thickness of the films: $t + t_{rough}$. With such thicknesses being fixed, the only parameters left free during the fit procedure were those of the Kramers Kronig – consistent oscillators describing the dielectric function ε of the film.

An excellent agreement between the best-fit spectra and the experimental ones is found for all the films (**Figure 1a** and **Annex A2**, **Figure S1**). The corresponding best-fit spectra obtained for the $ε_1$ and $ε_2$ of each Bi film are shown in **Figure 1b**. For the thinnest film ($t_{Bi}$ = 11 nm), a broad and moderately strong absorption band peaking at 0.9 eV and extending in the near infrared and visible is seen in the $ε_2$ spectrum. In relation with such absorption, $ε_1$ takes values close to 0 in the visible and ultraviolet together with relatively high positive values in the infrared. Upon increasing $t_{Bi}$ to 17 nm, the absorption band becomes markedly more intense and sharper, $ε_1$ taking large negative values in the visible and ultraviolet, and large positive values in the infrared. Upon increasing again $t_{Bi}$, the values of $ε_1$ and $ε_2$ converge asymptotically to those of bulk Bi, which are reached for the film with $t_{Bi}$ = 78 nm [1]. For bulk Bi, $ε_2$ reaches values of up to 120 at the absorption band maximum, $ε_1$ increases up to 100 in the infrared and decreases down to - 25 in the visible.

It is worth noting that the dielectric function of the Bi film with $t_{Bi}$ = 17 nm is already comparable with that of bulk Bi, with peak $ε_1$ and $ε_2$ values reaching 70% of the bulk ones. Therefore, the dielectric function of very thin Bi films, say with $t_{Bi}$ > 15 nm, is comparable with that of bulk Bi. In contrast, for thinner films (here, for $t_{Bi}$ = 11 nm) the dielectric function departs strongly from the bulk one. This trend correlates with the film structure, which is continuous for $t_{Bi}$ > 15 nm and discontinuous for $t_{Bi}$ = 11 nm.

**3. Structure of the Bi Films**

To demonstrate this correlation, we show the structure of selected films ($t_{Bi}$ = 11 nm, 21 nm and 78 nm) studied by SEM and AFM. The top-view SEM images presented in **Figure**

**2a** show that the film with $t_{Bi}$ = 11 nm presents a discontinuous near-percolation structure, where voids with a near 10 nm width separate clusters of densely packed/coalesced nanoparticles. The in-plane size of these clusters ranges from 50 to 150 nm. For larger $t_{Bi}$ values, the films present a continuous structure consisting of densely packed grains and few to no voids. Upon increasing $t_{Bi}$ from 21 nm to 78 nm, the in-plane size of the grains increases from 50-150 nm to 100-200 nm while the few remaining voids are filled.

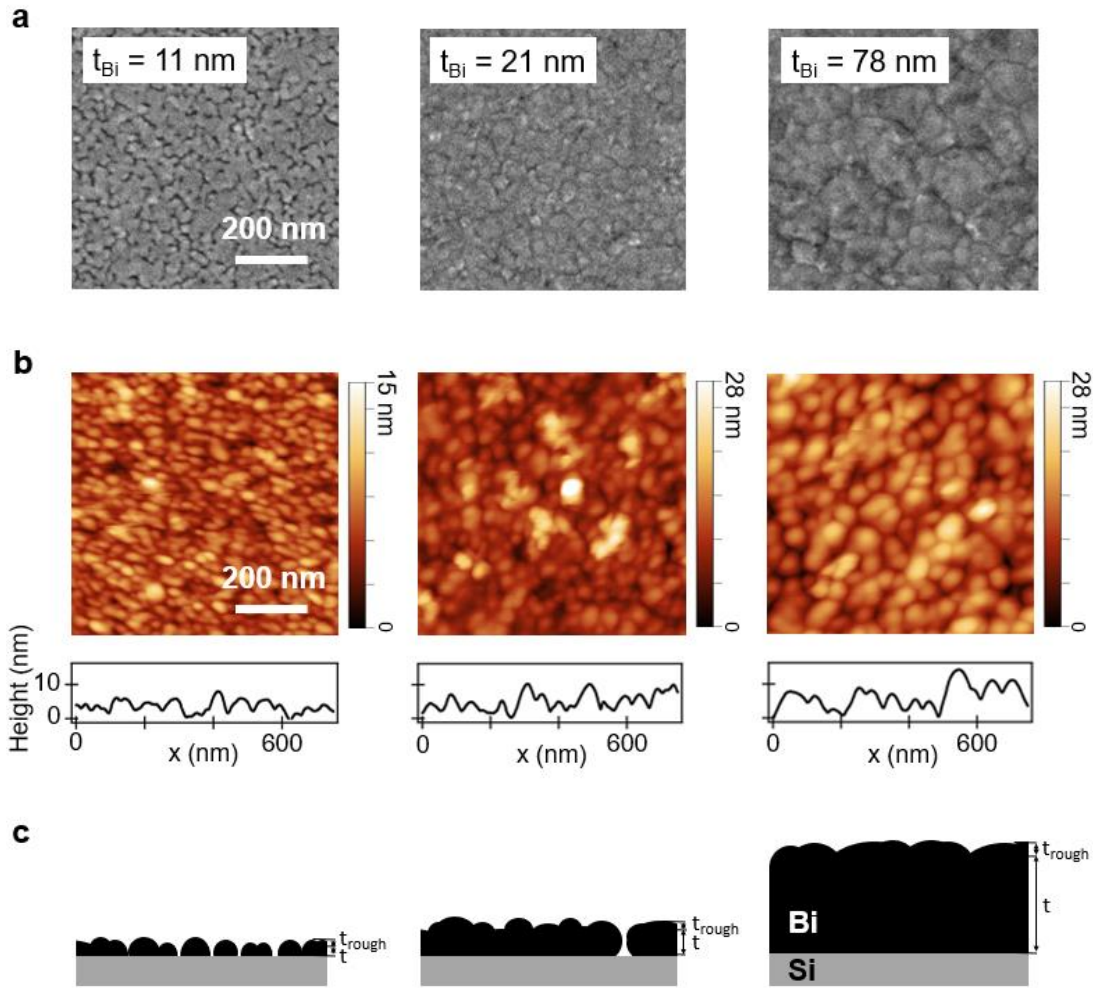

**Figure 2. Structure of selected Bi films.** (a) Top-view SEM images of the films with $t_{Bi}$ = 11 nm, 21 nm, and 78 nm. (b) AFM images and profiles of these films. (c) Cross-section schematic representation of the film structure. The film with $t_{Bi}$ = 11 nm has a discontinuous near-percolation structure. This structure is built from nanoparticles with a 20-50 nm in-plane size. They are arranged in clusters with a 50-150 nm in-plane size. The clusters are separated by voids with a near 10 nm width. Films with larger $t_{Bi}$ have a continuous structure with few/no voids. The corresponding thicknesses t and $t_{rough}$ are depicted for each film.

Complementary information is provided by the AFM images and profiles presented in **Figure 2b**. For the film with $t_{Bi}$ = 11 nm, they show that the in-plane size of the nanoparticles forming the clusters ranges from 20 to 50 nm. Note that AFM measurements do not reveal the full depth of the voids observed by SEM, because their width is too small compared with the tip size. For the same reason, they might underestimate the depth of the contact region between densely packed/coalesced nanoparticles. Therefore, combining SEM and AFM measurements is necessary to provide a full picture of the film structure, especially for discontinuous near-percolation films such as that with $t_{Bi}$ = 11 nm.

As summary, **Figure 2c** shows a cross-section schematic representation of the structure of the 3 films. For $t_{Bi}$ = 11 nm, the film has a discontinuous near-percolation structure. For $t_{Bi}$ = 21 nm, the film has a continuous structure with few voids. For $t_{Bi}$ = 78 nm, the film is also continuous yet without voids. These trends are a direct consequence of the growth mechanism of Bi on the surface-oxidized Si substrate, which follows a nucleation – growth – coalescence – percolation scheme as $t_{Bi}$ increases, in the same way as noble metals (**Annex A3**).

This growth mechanism impacts the optical properties of the film. Upon increasing $t_{Bi}$, the dielectric function of the Bi film gets closer to that of bulk Bi as the film becomes continuous and the density of voids decreases. It is especially remarkable that the strongest variation in $\varepsilon_1$ and $\varepsilon_2$ occurs when $t_{Bi}$ increases from 11 to 17 nm and the film structure turns from discontinuous to continuous with a few voids. The variation in $\varepsilon_1$ and $\varepsilon_2$ is smaller when $t_{Bi}$ increases above 17 nm and the few voids in the continuous film are gradually filled.

## 4. Discussion: Relation between the Structure and Optical Properties of the Bi Films

All the previous results point at a dominant effect of the Bi film discontinuity on the measured dielectric function when $t_{Bi}$ = 11 nm. In order to investigate the origin of such effect, finite difference time domain (FDTD) simulations of the optical properties of a discontinuous Bi film were performed. To simplify the problem while including the main structural features of such film, it was considered as a square array of densely packed truncated nanospheroids. We assumed that the material constituting these nanospheroids *has the same dielectric function as bulk Bi* taken from ref. 1. The nanospheroid height H was 17 nm, the nanospheroid diameter D was varied between 20 and 50 nm, and the separation gap G between nanospheroids was varied between 0 and 10 nm, in accordance with the geometrical film thickness t + $t_{rough}$ (**Annex A2**, **Table S1**), nanoparticle in-plane size and void width found for the film with $t_{Bi}$ = 11 nm. Further technical details about the FDTD simulations are given in **Annex A4** (**Figure S4**). In **Figure 3a** are shown the elementary cell used for the simulation with D = 50 nm and G = 10 nm and the corresponding electric field map at a photon energy of 1.5 eV. This map reveals mildly warm spots between the nanospheroids, which also present low internal and transmitted fields as a result of their strong absorption and dense packing.

FDTD simulations also provided the reflectance of the discontinuous film for the different values of D and G (**Annex A5**, **Figure S5**). The obtained reflectance values are much smaller than those of a continuous Bi film with the same layer thicknesses (t and $t_{rough}$) as the film with $t_{Bi}$ = 11 nm. This implies that the simulated *effective* dielectric function of the discontinuous film also presents much smaller values than the bulk one, as shown in **Figure 3b**. This trend is very similar to the one found for the measured dielectric function of the discontinuous Bi film with $t_{Bi}$ = 11 nm (**Figure 1b**).

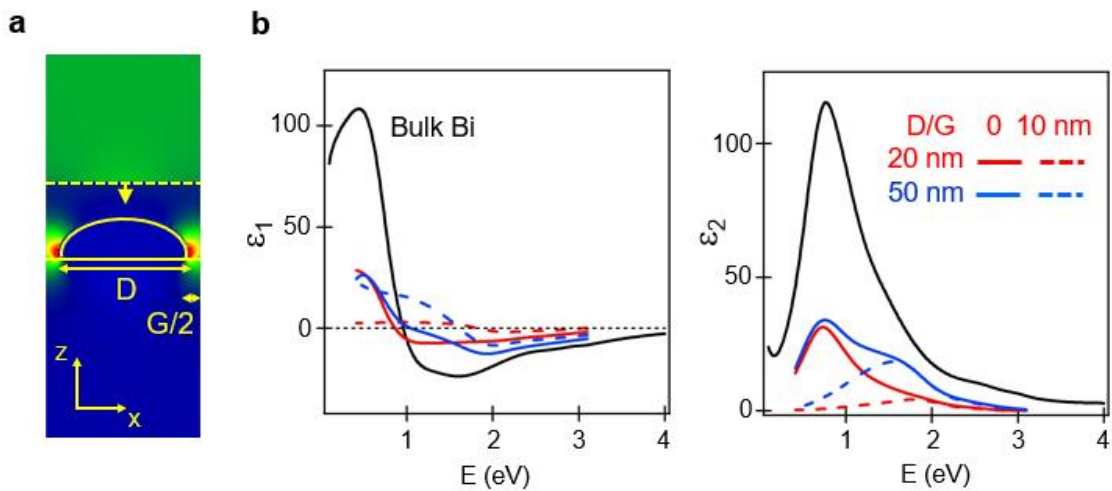

**Figure 3. FDTD simulations of the optical properties of discontinuous Bi thin films consisting of densely packed Bi nanospheroids.** (a) Example of elementary simulation cell seen in cross-section across the meridional plane of the nanospheroid, with a map of the x-polarized electric field amplitude at E = 1.5 eV (lowest value = 0, highest value = 1.5 relative to the incident field amplitude). In this example, the cell width is 60 nm, the nanospheroid diameter D = 50 nm and height H = 17 nm, the separation gap between nanospheroids is G = 10 nm. The incident light impinges downwards along the z axis and is x-polarized. (b) Simulated *effective* dielectric function (real part $\varepsilon_1$ and imaginary part $\varepsilon_2$) of discontinuous Bi films with different nanospheroid diameters D and separation gaps G (red and blue lines). The dielectric function of bulk Bi (black line) is shown for comparison. For any value of D and G, the discontinuous film presents much smaller values of $\varepsilon_1$ and $\varepsilon_2$ than bulk Bi.

Furthermore, simulations show that $\varepsilon_2$ peaks at a different photon energy (between 0.8 and 1.5 eV) depending on the values of D and G. Therefore, in a random array of nanospheroids characterized by a distribution of D and G, the $\varepsilon_2$ spectrum would display a broad band spreading in the near infrared and visible, as the one measured for the discontinuous Bi film with $t_{Bi} = 11$ nm (**Figure 1b**).

This leads us to propose that the dielectric function of the thinnest film studied here ($t_{Bi} = 11$ nm) is very different from that of bulk Bi mainly because of effective medium effects. The dielectric function measured for this film is an *effective* quantity resulting from the polarization of a Bi:void heterogeneous medium, where the Bi nanostructures are described by a dielectric function close to the one of bulk Bi. Such effective medium effects may also affect, yet more weakly, the dielectric function of thicker films still containing a few voids. This would explain their small difference with the dielectric function of bulk Bi.

In contrast, it seems that quantum electronic confinement does not have a dominant effect on the far infrared – to – ultraviolet dielectric function of even the thinnest film studied here, despite of the fact that it presents a discontinuous structure built from nanoparticles. This conclusion is in line with our previous work [5] in which we modeled satisfactorily the measured ultraviolet – visible plasmon – like resonances of flattened Bi nanoparticles using classical electrodynamic calculations based on the dielectric function of bulk Bi. Therefore, a continuous 10 nm – thick or even thinner Bi film might present a dielectric function comparable to the bulk one. A similar trend has been observed in the case of few – nm $Bi_2Se_3$ films based on a careful material fabrication and spectroscopic ellipsometry characterization [20, 21].

**5. Conclusion**

Summarizing, our results suggest that the outstanding optical properties of bulk Bi are shared by continuous Bi films down to the ultrathin film regime (thickness < 10 nm). This opens the way, for instance, to achieving a near total absorption of visible light with ultrathin and continuous Bi films. Also, they show that the dielectric function of bulk Bi can be used as input value for the rational design of flattened nanostructures such as nanocylinders or nanoflakes. Besides that, we also remark that the effective medium properties we put forward for the discontinuous Bi film are appealing for the fabrication of films with light trapping properties optimized by nanoscale design. Tailoring the film nanostructure [5, 22] enables tuning its effective optical dispersion in a broad range. This is useful to meet the requirements for an optimal light harvesting [23] in Bi nanostructured materials, in particular for photocatalytic systems based on near-percolation Bi nanostructures [24-25].

# ANNEX

## A1. Details about the Fabrication and Characterization of the Bi Films

*Fabrication and Structural Characterization.* The Bi films were grown by pulsed laser deposition on standard surface oxidized Si substrates. The number of laser pulses on a Bi target was controlled to tune the amount of deposited Bi and thus the *nominal thickness* $t_{Bi}$ of the films. Here, $t_{Bi}$ is an *equivalent thickness* defined as the geometrical thickness that a film would present if the deposited Bi atoms were perfectly arranged, i.e. if the film had a perfectly flat surface and a continuous structure with the atomic density $\rho_{Bi}$ of bulk Bi. It is determined from the areal density of Bi atoms $N_{Bi}$ measured by RBS, following the relation $t_{Bi} = N_{Bi}/\rho_{Bi}$.

The structure of the films was characterized by AFM and SEM. More details about the experimental procedures and the setups used are given in Ref. 1. The AFM characterization reveals surface roughness for all the films. Therefore, they can be considered as a bilayer stack, with a roughness layer of thickness $t_{rough}$ on a layer of thickness t. Here, $t_{rough}$ is taken equal to the average peak-to-valley roughness determined by AFM. To determine t, one writes the relation: $t_{Bi}$ = cov x (t + 0.5$t_{rough}$), where cov is the coverage of the substrate surface by the film. This relation expresses that the amount of deposited Bi determined by RBS is shared between the two layers in a proportion that depends on their thicknesses and on the substrate coverage. The SEM characterization provides the substrate coverage values: 80% for the thinnest film (discontinuous structure) and 100% for the others (continuous structure). One then obtains t from the relation: t = $t_{Bi}$/coverage – 0.5$t_{rough}$. Finally, from the obtained thicknesses, one determines the *geometrical film thickness*, which represents the distance between the substrate surface and the top of the film roughness: t + $t_{rough}$.

*Ellipsometry Characterization.* The ellipsometry spectra (ellipsometric angles $\Psi$ and $\Delta$) of the Bi films on the Si substrates were determined by spectroscopic ellipsometry at 2 angles of incidence (50º, 70º) in a range from 0.05 eV (far infrared) to 4 eV (near ultraviolet). Two ellipsometers (Woollam IR-VASE and Woollam VASE) were used to cover this spectral range. The back side of the Si substrates was roughened mechanically to avoid back reflected light to reach the detector. The ellipsometric characterization of the film with $t_{Bi}$ = 78 nm is detailed in ref. 1.

## A2. Ellipsometry Characterization of the Bi Films

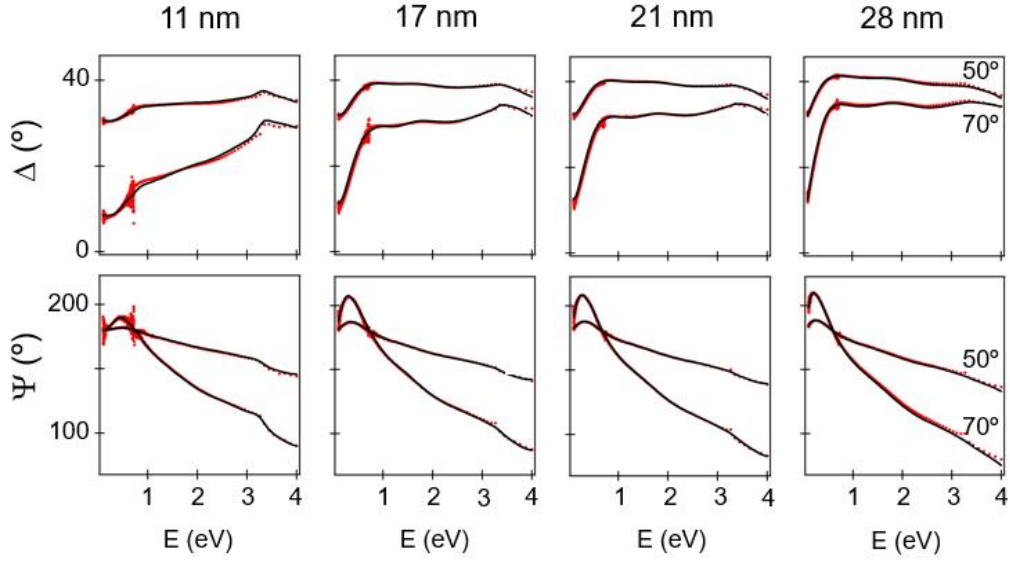

**Figure S1.** Spectra of the ellipsometric angles Ψ and Δ at the angles of incidence of 50° and 70°: experimental spectra (red dots) and the corresponding best-fit ones (black lines), for all the films. The spectra of the film with $t_{Bi}$ = 78 nm are shown in ref. 1.

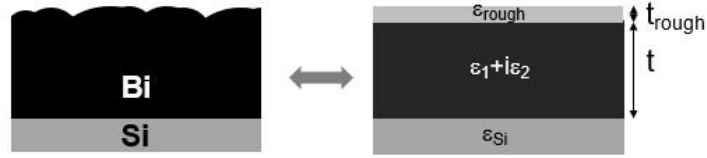

**Figure S2.** Bilayer structure used to model the Bi films for ellipsometry fitting.

| $t_{Bi}$ (nm) | $t_{rough}$ (nm) | t (nm) | t + $t_{rough}$ (nm) |
|---|---|---|---|
| 11 | 6 | 11 | 17 |
| 17 | 7 | 13 | 20 |
| 21 | 8 | 17 | 25 |
| 28 | 8 | 24 | 32 |
| 78 | 8 | 74 | 82 |

**Table S1.** Nominal thickness $t_{Bi}$ of the films, layer thicknesses $t_{rough}$ and t used for the ellipsometry analysis, and geometrical thickness of the films, t + $t_{rough}$. The methods used to determine thicknesses are explained in **Annex A1**. Note that $t_{Bi}$ is the geometrical thickness that a film would present if its Bi atoms were perfectly arranged, i.e. if it had a perfectly flat surface and a continuous structure with the atomic density of bulk Bi. However, in this work the films are not perfectly flat and the thinnest one has a discontinuous structure. Therefore, for these films $t_{Bi}$ is smaller than the geometrical film thickness, t + $t_{rough}$.

## A3. Growth Mechanism of the Bi Films

In our deposition conditions, nucleation and growth (steps 1 and 2) occur for $t_{Bi}$ up to 1-2 nm with the formation of truncated spheroidal nanoparticles [5]. For larger $t_{Bi}$ up a few nm, coalescence events (step 3) make the nanoparticles grow laterally [5]. Percolation is reached for $t_{Bi}$ above 10 nm for which the deposit consists of a discontinuous layer (step 4). In line with that, the film with $t_{Bi}$ = 11 nm studied in this work displays a near-percolation structure. Upon further increasing $t_{Bi}$ the voids within the discontinuous layer get filled by Bi to promote the formation of a continuous layer. For $t_{Bi}$ > 15 nm, a continuous layer with few voids is already formed. Further Bi deposition leads to a filling of the remaining voids (step 5) and to a vertical growth that leads to an increase in the continuous layer thickness (step 6). Even when consisting of a continuous layer with $t_{Bi}$ > 30 nm, the Bi deposit presents a small roughness reminiscent from the initial stages of growth.

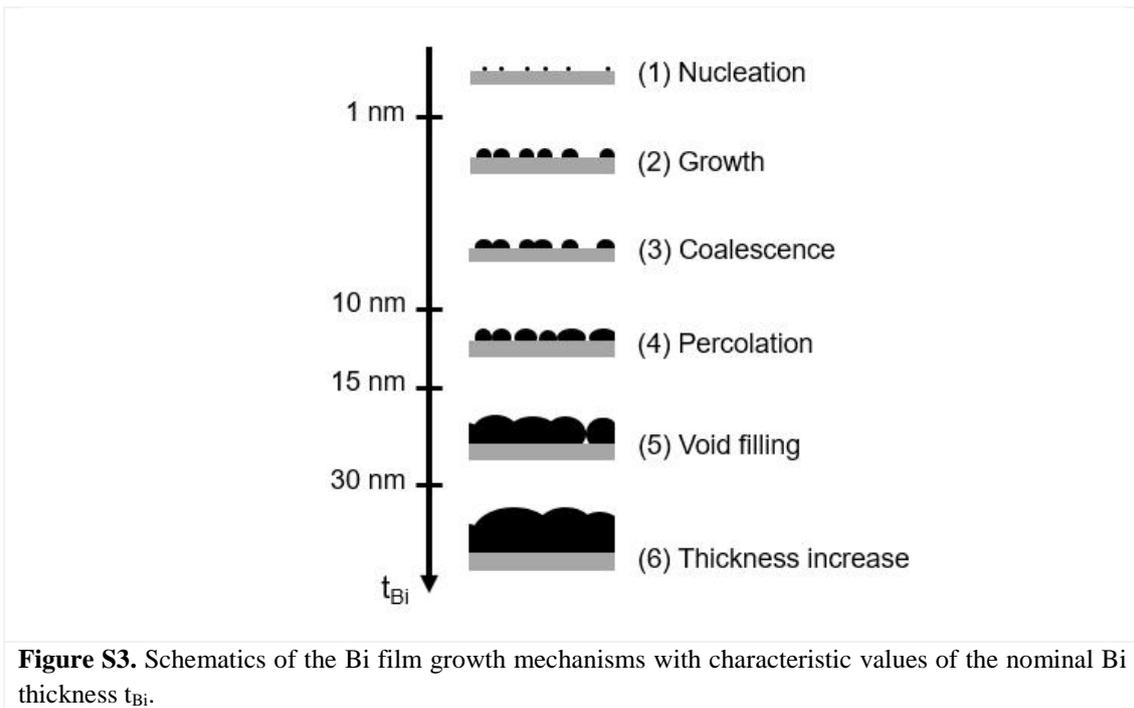

**Figure S3.** Schematics of the Bi film growth mechanisms with characteristic values of the nominal Bi thickness $t_{Bi}$.

## A4. Details about the FDTD Simulations of Discontinuous Bi Films

3D-FDTD simulations were performed with the OptiFDTD32 software. A parallelepipedal elementary cell, with the geometry shown in **Figures S4a** and **S4b**, was used.

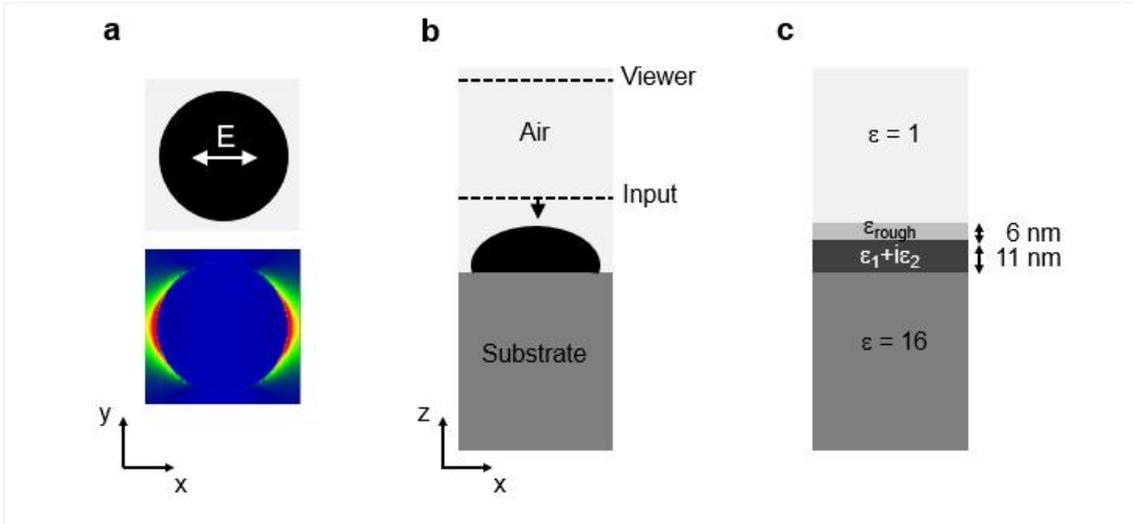

**Figure S4.** (a) Top view of the elementary cell used for the simulation, with the corresponding x-polarized electric field amplitude map in the equatorial plane of the Bi nanospheroid, for D = 50 nm, H = 17 nm and G = 10 nm. E is the incident electric field. (b) Cross-section of the elementary cell, showing the location of the input plane (incident field) and of the viewer plane (where the reflectance spectrum is measured). (c) Bilayer structure used to determine the effective dielectric function of the Bi discontinuous film from the FDTD – calculated reflectance spectra.

*Cell Elements.* In this cell, a Bi nanospheroid was standing on a substrate with a dielectric constant of 16 (comparable with that of Si) and was surrounded by a medium with a dielectric constant of 1 (air). Its dielectric function was described as the sum of Kramers Kronig consistent oscillators with the parameters given for bulk Bi in ref. 1. The truncated nanospheroid height H was set to 17 nm, to match with the geometrical thickness (t + $t_{rough}$ = 11 + 6 nm) of the film with $t_{Bi}$ = 11 nm. Its diameter D was varied between 20 and 50 nm.

*Cell Dimensions and Boundary Conditions.* The height of the elementary cell was 150 nm, and its width W was adjusted to fulfil W = G + D, G being the separation gap between nanospheroids in the square array, varied between 0 and 10 nm. To generate such array, periodic boundary conditions were applied on the ±x and ±y boundaries. To account for a semi-infinite incident medium and substrate, perfect matching layers were applied to the ±z boundaries.

***Wave Parameters and Reflectance Simulations.*** The incident electric field was x-polarized and incident from the air medium, downwards along the z axis, i.e. at normal incidence. Reflectance spectra were calculated by inverse Fourier transform of a light pulse, analyzed in the viewer plane located ~ 60 nm above the nanospheroid. This distance is sufficiently large so that the viewer plane lies in the far-field region. A mesh size of 0.5 nm was used for all the calculations and enabled obtaining converged reflectance spectra.

***Effective Dielectric Function.*** To determine the effective dielectric function $\varepsilon = \varepsilon_1 + i\varepsilon_2$ of the discontinuous Bi film formed by the square array of nanospheroids, its FDTD – calculated reflectance spectrum was fitted assuming a bilayer structure. This structure is shown in **Figure S4c**. $\varepsilon$ consisted of a sum of Kramers Kronig - consistent oscillators whose parameters were left free during the fit. $\varepsilon_{rough}$ was described by a Bruggeman model mixing $\varepsilon$ and the dielectric function of air with equal weight. The thicknesses of the bottom and top layer were set to 11 and 6 nm, respectively, to match with the layer thicknesses t and $t_{rough}$ used to model the film with $t_{Bi} = 11$ nm and so that the corresponding geometric film thickness equals the nanospheroid height H.

## A5. FDTD Reflectance Spectra of Discontinuous Bi Films

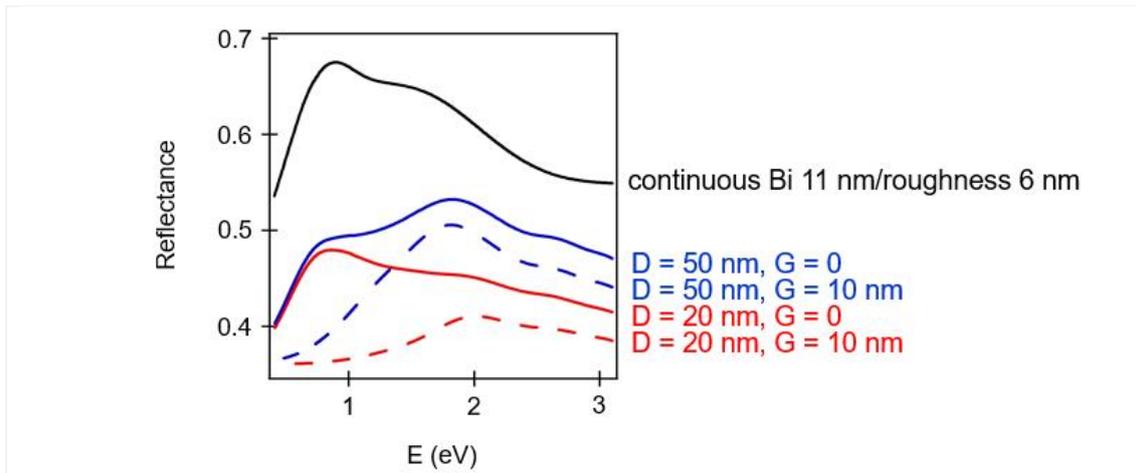

**Figure S5.** FDTD reflectance spectra for discontinuous Bi films consisting of a square array of Bi truncated nanospheroids, with different diameter D and separation gap G, and a height H = 17 nm. These spectra are compared with that of a continuous Bi film with the same layer thicknesses (t = 11 nm, $t_{rough}$ = 6 nm) as the film with $t_{Bi}$ = 11 nm.

**ACKNOWLEDGEMENTS**

The authors thank Drs. J. Wojcik and P. Mascher from McMaster University (Canada) for performing the far infrared ellipsometry measurements. They also thank Dr. T.A. Ezquerra from the Instituto de Estructura de la Materia, IEM, CSIC (Spain) for providing access to the AFM facility. This work was partially funded by the Spanish Ministry of Science, Innovation and Universities [projects MINECO/FEDER TEC2015-69916-C2-1-R and MICINN RTI2018-096498-B-I00].